# Near-field plates and the near zone of metasurfaces

R. Merlin

*The Harrison M. Randall Laboratory of Physics, University of Michigan,
Ann Arbor, Michigan 48109-1040, USA*

**Abstract:** A brief, tutorial account is given of the differences between the near and far regions of the electromagnetic field emphasizing the source-dependent behavior of the former and the universal properties of the latter. Field patterns of near-field plates, that is, metasurfaces used for subwavelength applications, are discussed in some detail. Examples are given of fields that decay away from the plates in an exponential manner, a ubiquitous feature of many interface problems, and metasurfaces for which the decay is not exponential, but algebraic. It is also shown that a properly designed system of two parallel near-field plates can produce fields that exhibit pseudo minima, which are potentially useful for near-field tweezer-like applications.

## 1. Introduction

Metasurfaces are artificial two-dimensional structures used to generate a desired electromagnetic (EM) field pattern or modify an incoming wave to obtain a predetermined result [1]. Most of the EM applications of metasurfaces, from microwave receivers and transmitters [2] to reconfigurable devices [3] and flat optical lenses [4], pertain to the radiation (or far) zone. Metasurfaces whose primary function involves the near field are known as near-field plates (NFPs) [5,6]. Since the near zone encodes detailed information about the sources, ignoring restrictions imposed by the standard diffraction limit [7] and, moreover, because it allows one to separate the electric from the magnetic field, the interest in NFPs centers primarily on subwavelength focusing [8] and wireless power transfer [9].

In this work, we present an abridged review of the near field properties of metasurfaces, emphasizing the forms of decay of the EM field (although the elastic field is not considered here, many of the results apply also to acoustic metasurfaces [10]). Other than the ever present exponential decay, commonly associated with interface phenomena [11], we show cases where the decay of the near field is algebraic in nature. We also introduce an arrangement of a pair of metasurfaces and describe its potential use as EM tweezers.

## 2. The near field and evanescent waves

### 2.1 Localized charges and currents

Textbooks tell us that the EM field of a confined distribution of moving charges behaves very differently in regions that are close to and far from the charges, with the length scale determined by the wavelength of the radiation, $\lambda$ [12,13]. Differentiation between the near and far behaviors appears already in the expressions for the fields resulting from the motion of a point charge. The corresponding Liénard–Wiechert potentials involve the sum of two terms: (*i*) the near-field contribution, which is associated with the static fields and does not contain the acceleration of the sources, and (*ii*) the radiation or far-field term, which vanishes as the acceleration goes to zero. The separation is also apparent in the expressions for the EM field of a sinusoidally, time-varying electric (magnetic) dipole where the electric (magnetic) field dominates in the near zone whereas, far away from the dipole, the electric and magnetic field are of the same magnitude (Gaussian units), and both decay inversely proportional to the distance. For various reasons, and leaving aside the question of the sources needed to generate a particular field, it is useful to frame the broad near-zone vs. far-zone discussion around the behavior of the EM field in vacuum. For fields with time dependence given by $e^{-i\omega t}$ ($\omega$ is the angular frequency), the empty-space potentials in the Lorenz gauge satisfy the homogeneous Helmholtz equation

$$\left(\nabla^2 + \omega^2/c^2\right)F = 0 \tag{1}$$

where $F$ is the electrostatic potential $\Phi$ or a cartesian component of the potential vector $\mathbf{A}$, and $c$ is the speed of light, with $\nabla \cdot \mathbf{A} - i\omega c^{-2}\Phi = 0$. For localized sources, it is convenient to expand $F$ in terms of outgoing spherical waves,

$$F = \sum_{lm} A_{lm} h_l^{(1)}(kr) Y_{lm}(\theta,\varphi) \ , \tag{2}$$

themselves solutions of Eq. (1). Here, $r$ is the radial distance, $\theta$ and $\varphi$ are the polar and azimuthal angles, $k = \omega/c$ is the wavevector and $A_{lm}$ are expansion coefficients; $h_l^{(1)}$ and $Y_{lm}$ are, respectively, a spherical Hankel function of the first kind and order $l$ and a spherical harmonic. Since $h_l^{(1)} \to e^{ikr}/r$ for $kr \gg 1$, regardless of $l$, the far-field $1/r$ decay behavior is universal. Instead, the near-field properties depend on the particulars of the distribution. Consider sources confined to a region of space of dimensions $d \ll \lambda$. Then, for $d \ll r \ll \lambda$, $h_l^{(1)} \sim r^{-(l+1)}$ and, since $\nabla^2 \left[ r^{-(l+1)} Y_{lm}(\theta, \varphi) \right] = 0$, we get $\nabla^2 F \approx 0$. This is to be expected given that the Helmholtz equation becomes Laplace's equation in the limit $c \to \infty$, when retardation can be ignored (since magnetism is a relativistic effect, care must be exerted when taking this limit for **A**). To lowest order, we have $\Phi \sim 1/r$ and $A_i \sim 1/r^2$, which correspond, respectively, to the fields of a static electric monopole and magnetic dipole. We note that, unlike the Liénard–Wiechert expressions, the spherical-wave expansion does not split into a separate sum of near- and far-zone terms. Rather, each term in Eq. (2) gives the corresponding near- and far-field expressions in two separate limits.

## 2.2 Sources behind a plane: Cylindrical and Cartesian waves, and exponential decay

The above considerations are not applicable to extended sources of dimensions $\gtrsim \lambda$ or, for that matter, to metasurfaces, which divide space into two halves. Let us assume that all the sources are in the half-space defined by $z < 0$ and expand the empty half-space potentials in terms of the complete set of solutions of Helmholtz equation in cylindrical coordinates

$$F(\rho, \varphi, z) = \sum_n e^{in\varphi} \int B_n(q) J_n(q\rho) e^{i\kappa(q)z} dq \qquad (3)$$

where $\rho = \sqrt{r^2 - z^2}$, $\kappa^2 = k^2 - q^2$, $J_n$ is a Bessel function of order $n$ and $B_n$ are the parameters of the expansion. Notice that $B_n = B_0 \delta_{n0}$ for azimuthally symmetric fields like, e.g., the axicon [14]. Eq. (3) divides into exponentially-decaying components for which $q > k$, and travelling waves, commonly known as Bessel beams [15], with real $\kappa < k$. This allows for a clear separation between the near and far fields, associated primarily with the evanescent ($\kappa = i|\kappa|$) and travelling components, respectively. Using the orthogonality condition for Bessel functions, we get

$$B_m(q) = \frac{q}{2\pi} \int F(\rho, \varphi, 0) e^{-im\varphi} J_m(q\rho) \rho \, d\rho \, d\varphi \qquad , \qquad (4)$$

which, together with Eq. (3), establishes an exact relationship between values of the field at two parallel planes (two values of $z$).

Consider now the equivalent Cartesian coordinate approach [16,17]. Let $F(x, y, z_0)$ be the potential field in the plane $z = z_0$. The angular spectrum in this plane is defined as the Fourier transform

$$\alpha(q_x, q_y, z_0) = \frac{1}{4\pi^2} \iint F(x, y, z_0) e^{i(q_x x + q_y y)} dx dy . \tag{5}$$

As before, we assume that all the sources are in the half-space $z < 0$. Then, the field at any point in the source-free half-space is given by

$$F(x, y, z) = \int \alpha(q_x, q_y, z_0) e^{i\left[q_x x + q_y y + \kappa(z - z_0)\right]} dq_x dq_y \tag{6}$$

where

$$\kappa = \begin{cases} i\left|(q_x^2 + q_y^2 - k^2)^{1/2}\right| & q_x^2 + q_y^2 \geq k^2 \\ \left|(k^2 - q_x^2 - q_y^2)^{1/2}\right| & q_x^2 + q_y^2 < k^2 \end{cases} . \tag{7}$$

Note that, by construction, $(\nabla^2 + \omega^2/c^2)F = 0$. Like its cylindrical counterpart, Eqs. (3-4), this expression can be used (*i*) to calculate the field propagation in the forward direction or, by back propagation, (*ii*) to infer the field at $z = z_0$ that will produce a desired field pattern further ahead. The latter approach was applied to the design of a NFP [5] mimicking the behavior of a negative-index slab [18] to attain focusing beyond the standard diffraction limit; one of the first realizations of a metasurface [6].

As for the cylindrical-wave representation, the (cartesian) angular decomposition allows for a sharp separation between the near- and the far-field depending on whether $\kappa$ is purely imaginary or real [19]. This, however, should not be construed to imply that the decay of the near-field component of $F$ is always exponential, for planar geometries or otherwise (recall that the near field of a localized charge distribution decays algebraically). Several examples of non-evanescent decay of the metasurface near field are given below. Eq. (6) shows that a sufficient condition for exponential decay to occur is to have a sharp peak in the angular spectrum at a wavevector $\mathbf{q}_0$ of modulus $q_0 > k$ so that $F \sim e^{-(q_0^2 - k^2)^{1/2} z}$. This is precisely the condition met by important interface phenomena, such as total internal reflection, surface polaritons [20] and focusing by a negative-index slab [18]. Similar considerations apply to cylindrical fields for which $B_m(q) \propto \delta(q - q_0)$ with $q_0 > k$; see Eqs. (3-4).

## 3. One-dimensional metasurfaces and near-field plates

From now on, and for simplicity, we focus the attention on problems for which $\partial F / \partial x \equiv 0$. In this case, the most general outgoing-wave solution to Helmholtz' equation is of the form

$$F = \sum_n C_n H_n^{(1)}(k\eta) e^{in\beta} \qquad (8)$$

where $\eta^2 = y^2 + z^2$ and $\tan \beta = y/z$. $H_n^{(1)}$ is a Hankel function of the first kind and order $n$, and $C_n$ are constants. Since $\eta = 0$ is the only singularity of the Hankel functions, this expansion accounts for the field on the source-free side of a one-dimensional metasurface placed at arbitrary $z > 0$. Once again, the behavior of the (two-dimensional) far field is universal since $H_n^{(1)}(k\eta) \to e^{ik\eta} / \eta^{1/2}$ for $k\eta \gg 1$, while the near field properties depend on the specifics of the source distribution. Replacing the Hankel functions by their small argument limit, we get

$$\lim_{k\eta \to 0} F = \frac{2iC_0}{\pi} \ln(k\eta/2) - \frac{i}{\pi} \sum_n C_n \Gamma(n) (2/k\eta)^n e^{in\beta} \qquad . \qquad (9)$$

This expression is formally identical to the general solution of the two-dimensional Laplace's equation in polar coordinates that is consistent with $F \to 0$ at infinity. With $F$ becoming a har-monic function (that is, a function that satisfies $\nabla^2 F = 0$ in two dimensions) and, after some rearrangement of terms, the limit $k\eta \to 0$ leads to the multipole expansion of the static electric ($\Phi$) and magnetic (**A**) potentials (we observe once more that caution must be applied when calculating **A** since the magnetic field vanishes for $c \to \infty$).

Fig. 1 shows results for

$$F = \sum_{n=0}^{N} \frac{H_n^{(1)}(k\eta)}{H_n^{(1)}(kR)} \cos n\beta \quad . \tag{10}$$

At large $N$, this expression gives a field that is sharply peaked at $\beta = 0$ in the circle $\eta = R$. The contour plot, Fig. 1A, and Fig. 1B ($y = 0$) reveal three distinctive regions with the subwavelength scale $R$ setting the boundary between the two that belong to the near field. For $y = 0$ and $z \ll R \ll \lambda$, the behavior of $F$ is determined by $H_N^{(1)}$, the leading term of which is $z^{-N}$ while, in the intermediate near-zone region $R \ll z \ll \lambda$, the dominant term is $H_0^{(1)} \sim \ln kz$. As expected, $F \propto z^{-1/2}$ in the far field ($kz \gg 1$). Also important is that, for $z \ll R$, $F$ shows a peak of width $\approx z/N$ centered at $y = 0$ (Fig. 1A covers too large a range to notice the $z$-dependence of the width).

Returning to the angular spectrum representation, consider a situation where $F(y,0)$ is localized mainly in a segment of length $\ll \lambda$, as in the above example and, more generally, for a generic NFP. We can then ignore contributions to $\alpha(q,0)$ from all, but spatial frequencies $\gg k$, so that the near field can be approximated by the harmonic expression

$$F(y,z) \approx \int \alpha(q,0) e^{iqy} e^{-|q|z} dq \quad . \tag{11}$$

It is well known that arbitrary functions of the form $f(iy \pm z)$ are solutions to Laplace's equation in two dimensions. Thus, the near field associated with a subwavelength-localized $F(y,0)$ is a harmonic function that is analytic in the half-space $z > 0$ and decays with increasing $z$.

## 4. Modulated grating and single-aperture near-field plates

Figure 2 shows calculations using Eq. (11) for

$$F(y,0) = \frac{e^{iq_0 y}}{y^2 + L^2} \quad ; \tag{12}$$

the corresponding angular spectrum is $\alpha(q,0) = e^{-L|q-q_0|}/2L$. The modulated-grating profile mirrors the behavior of the EM field at the exit side of a negative-index slab [5,21]. If $q_0 \gg k$, near identical results are obtained using the exact expression, Eq. (6). We emphasize that the plots show only the near field. Because of the dominant peak of $\alpha(q,0)$ at $q = q_0$, $F$ decays first exponentially, turning later into a $1/z$ dependence for $z \gg q_0^{-1}$. Not shown in Fig. 2 is the far field, which manifests itself for $z \gg k^{-1}$ and decays $\sim 1/z^{1/2}$ for $y = 0$, as expected. For completeness, we give below the explicit expression of the near field in terms of harmonic functions:

$$F(y,z) = \left[ \frac{1}{(L+iy+z)} - \frac{1}{(L+iy-z)} \right] e^{-q_0 L} + \frac{2 e^{q_0(iy-z)} L}{(L+iy-z)(L-iy+z)} \quad . \tag{13}$$

From here, one can show that, as the distance $z$ from the metasurface increases, the width of $|F(y,z)|$ decreases reaching a minimum value $\sim q_0^{-1}$ at $z = L$. This closely simulates the behavior of subwavelength focusing by a negative-index slab [18]. Because the field magnitude decreases with increasing $z$, the focus is a saddle point, something that could have been anticipated since harmonic functions do not have absolute maxima or minima [22].

Fig. 3 shows an example of near-field focusing without exponential decay. The calculations are for a single-aperture NFP with

$$F(y,0) = \pi/2 - \arctan\frac{y^2 - \Delta^2}{L^2} \quad . \tag{14}$$

This expression gives a Lorentzian-like peak of half-width $L$ for $L \gg \Delta$ whereas, for $L \ll \Delta$, $F \approx 0$ except for $|y| < \Delta$ where it is nearly constant. The corresponding angular spectra are roughly of the form $e^{-|q|L}$ and $\sin(q\Delta)/q$, respectively. Using these approximations and Eq. (11), we get

$$F(y,z) \approx \begin{cases} \dfrac{L+z}{y^2 + (L+z)^2} & (L \gg \Delta) \\ \dfrac{1}{2}\left(\tan^{-1}\dfrac{\Delta+y}{z} + \tan^{-1}\dfrac{\Delta-y}{z}\right) & (L \ll \Delta) \end{cases}. \quad (15)$$

In the limit $L \gg \Delta$ ($L \ll \Delta$), and for $z \ll L$ ($\Delta$), $F(0,z)$ is nearly constant while $F(y,z)$ exhibits a peak at $y = 0$ of half-width $\approx L$ ($\Delta$). At large values of $z$ (but $z \ll \lambda$), $F \propto 1/z$. The case $\Delta = 10$, $L = 1$ is illustrated in Fig. 3. As for the calculations of Fig. 2, we underline the fact that these plots depict only the near field. The results show that neither the amplitude (Fig. 3B) nor the width of the field vary much in the range $z \ll \Delta$. This makes the single-aperture NFP a promising candidate for subwavelength focusing and power-transfer applications, with the drawback that the high resolution is restricted to distances from the NFP on the order of the resolution itself (the same limitation as in evanescent-wave focusing [23]).

## 5. Pair of parallel near-field plates

As discussed earlier, at distances $z \ll \lambda$, the near-field of a metasurface is well described by a harmonic function that decays away from the plate. For a pair of parallel plates, however, the decay requirement does not longer apply and, thus, an arbitrary singularity-free harmonic function represents a physically realizable near field. This offers a path for the development of near-field devices that could meet the needs of particular applications. Of interest here is the design of near-field structures that can be used to manipulate subwavelength objects by means of radiation pressure, like optical tweezers. Even though harmonic functions such as $F$ do not possess maxima or minima [22], the saddle points make it possible to devise an intensity pattern that comes close to providing a confinement potential for the intensity. Fig. 4 shows one such an example. The three-dimensional plot gives $\text{Re}^2[F(y,z)]$, which is proportional to the intensity, for

$$F = \frac{e^{q_0(iy-z)} - 1}{iy - z} + \frac{e^{q_0(iy+z)} - 1}{iy + z} - V_0 \quad , \tag{16}$$

with parameters $q_0 = 4$ and $V_0 = 8$. The latter value was chosen so that there is a saddle point at $F \approx 0$. The depression around the origin is not a true minimum for it is possible to descend towards regions of lower intensity using the canyon-like features along the planes $x = \pm 1$. Nevertheless, the dip can be used to trap a particle with a refractive index smaller than that of the surrounding medium provided its size is large enough to prevent escape through the canyons.

## 6. Conclusions

We showed that, unlike the far field, the EM field close to a metasurface exhibits a variety of dependencies with distance, from that of evanescent waves, often (and wrongly) viewed as the hallmark of the near field, to various forms of algebraic decay. We also demonstrated that fields between two parallel near-field plates can have pseudo minima, a property that is potentially useful for applications as near-field EM tweezers.

# Figure Captions

FIGURE 1. Near and far field of a one-dimensional metasurface; see Eq. (10). (A) Contour plot of the normalized field, $\text{Re}[F(y,z)]/|F(0,z)|$, for $N = 20$, $R = 10$ and $\lambda = 10^6$. (B) $\ln|F(0,z)|$ vs $z$ (blue curve). Gray curves are asymptotes. For small and intermediate distances from the metasurface, $F \propto z^{-N}$ and $F \propto \ln z$ ($10^2 \lesssim z \lesssim 10^5$). These near-field forms correspond, respectively, to the small argument limit of $H_N^{(1)}$ and $H_0^{(1)}$. In the far field, $z \gtrsim 10^5$, $F \propto 1/\sqrt{z}$.

FIGURE 2. Near field of the modulated-grating metasurface; see Eq. (12). (A) Contour plot of the normalized intensity, $\text{Re}^2[F(y,z)]/\text{Re}^2[F(0,z)]$, for $q_0 = 7$ and $L = 2.5$. (B) $\ln|F(0,z)|$ vs. $z$ (blue curve). Gray curves are asymptotes. For small (large) $z$, $F$ decays exponentially ($F \propto 1/z$).

FIGURE 3. Near field of the single-aperture metasurface; see Eq. (14). (A) Contour plot of $\text{Re}^2[F(y,z)]/\text{Re}^2[F(0,z)]$ for $\Delta = 1$ and $L = 10$. (B) $\ln|F(0,z)|$ vs $z$ (blue curve). Gray curves are asymptotes. For small (large) $z$, $F$ is nearly constant ($F \propto 1/z$).

FIGURE 4. Near field of a system of two-metasurfaces; see Eq. (16). The 3D plot shows the square of the field, $\text{Re}^2[F(y,z)]$.

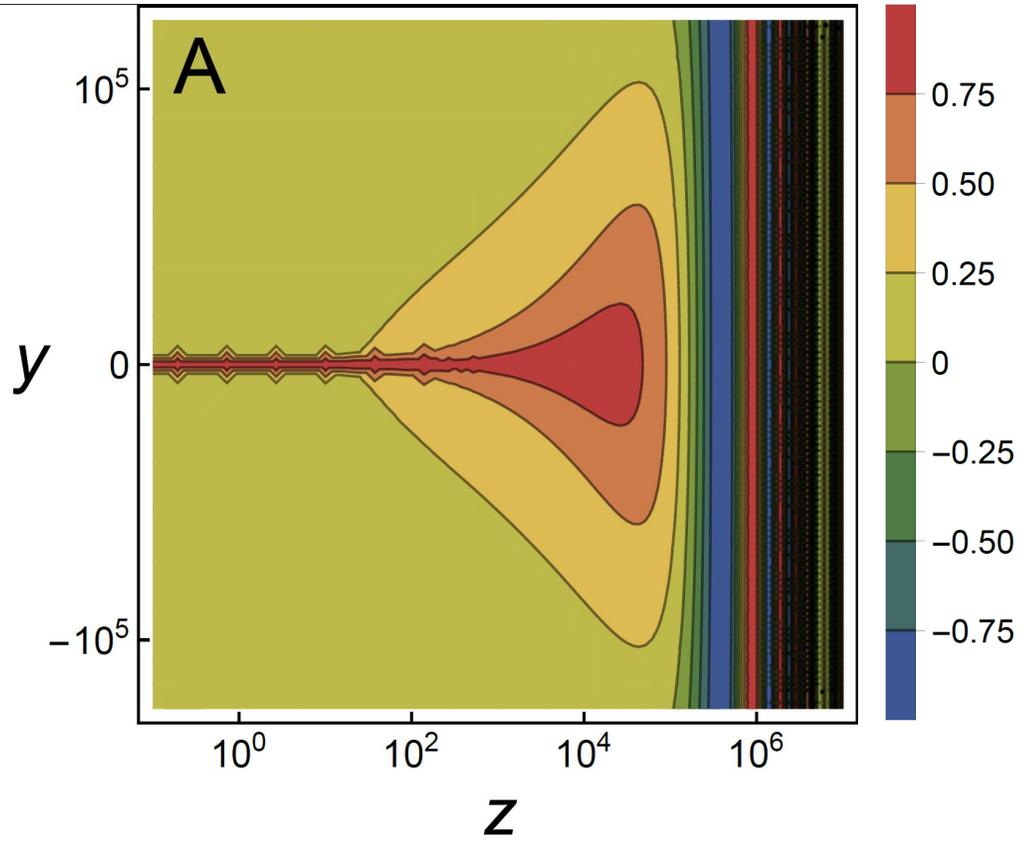

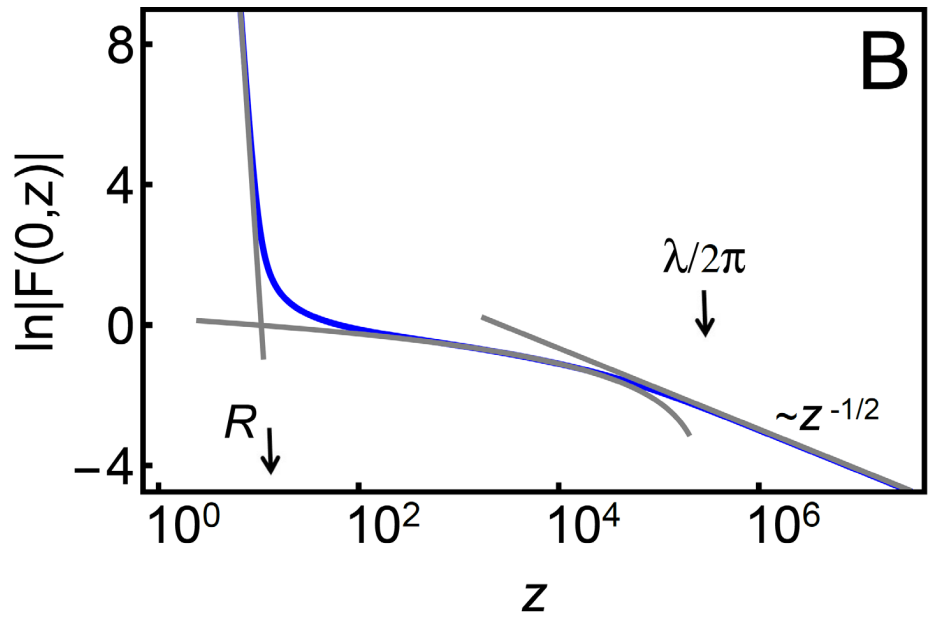

FIGURE 1

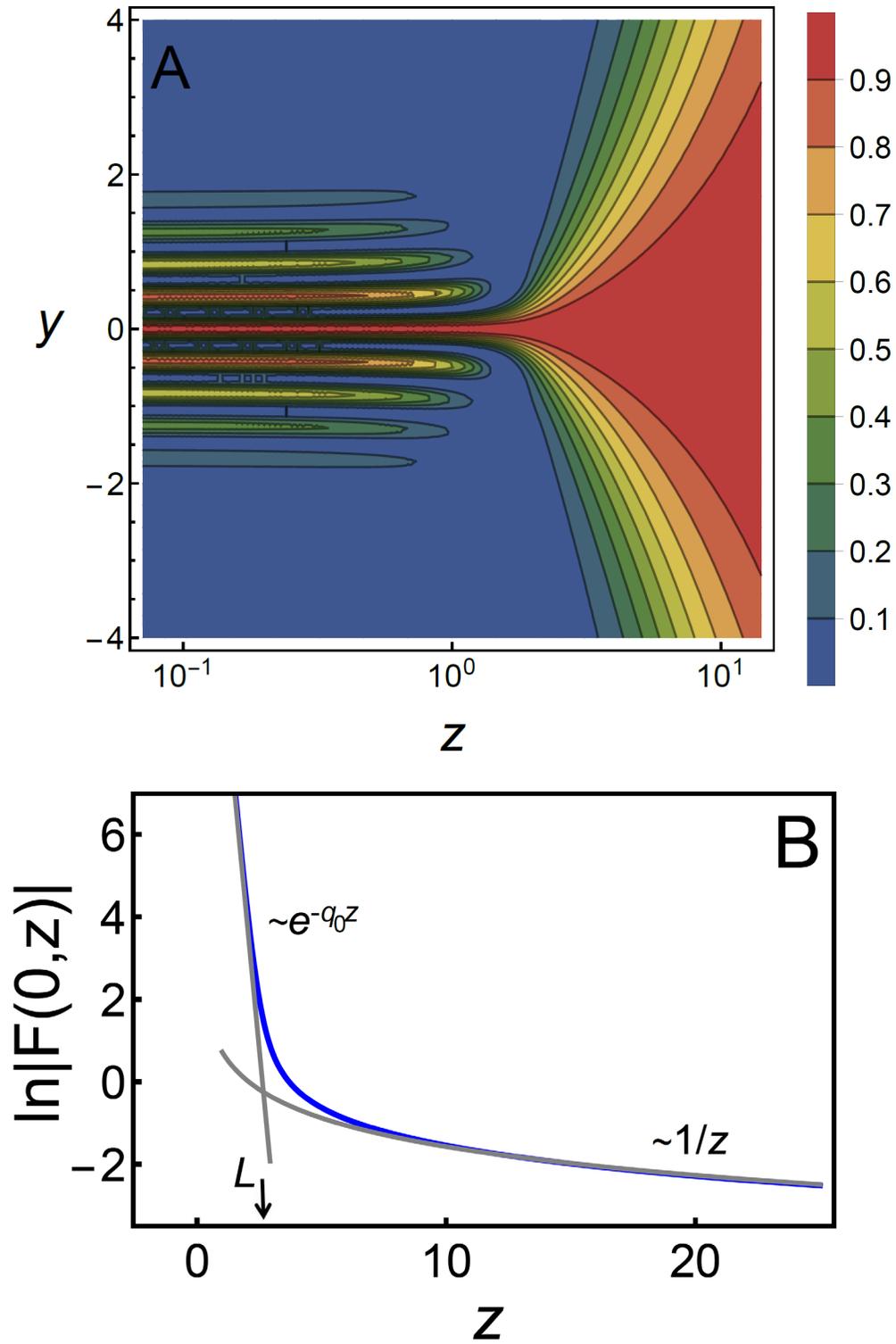

FIGURE 2

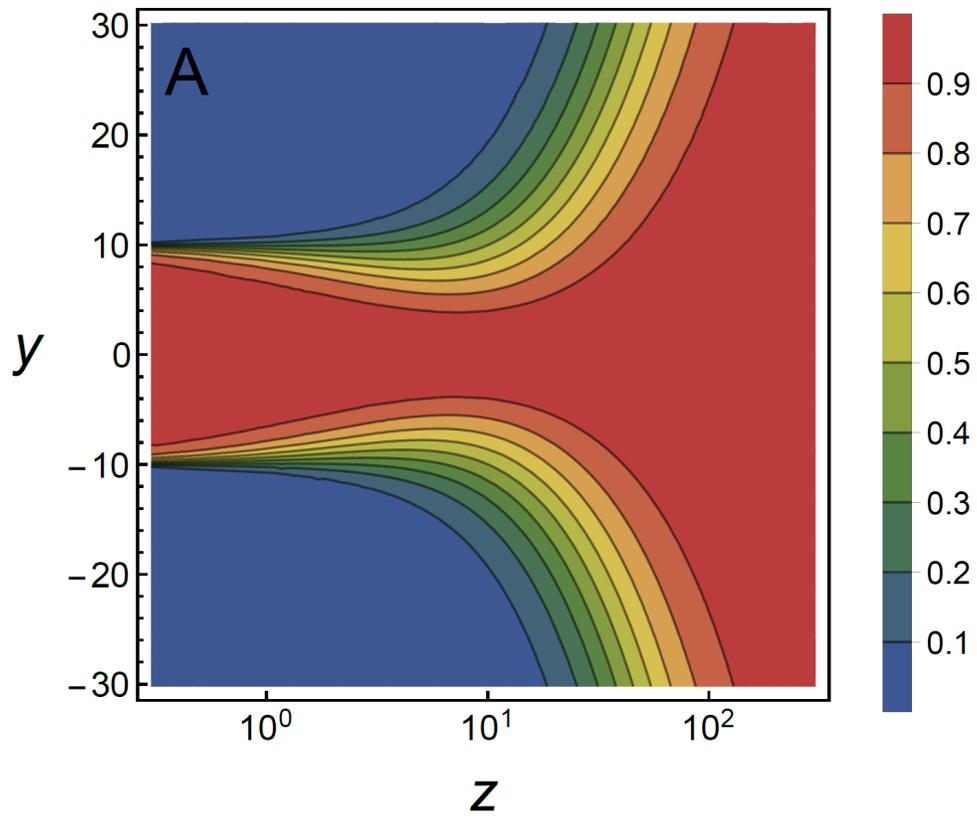

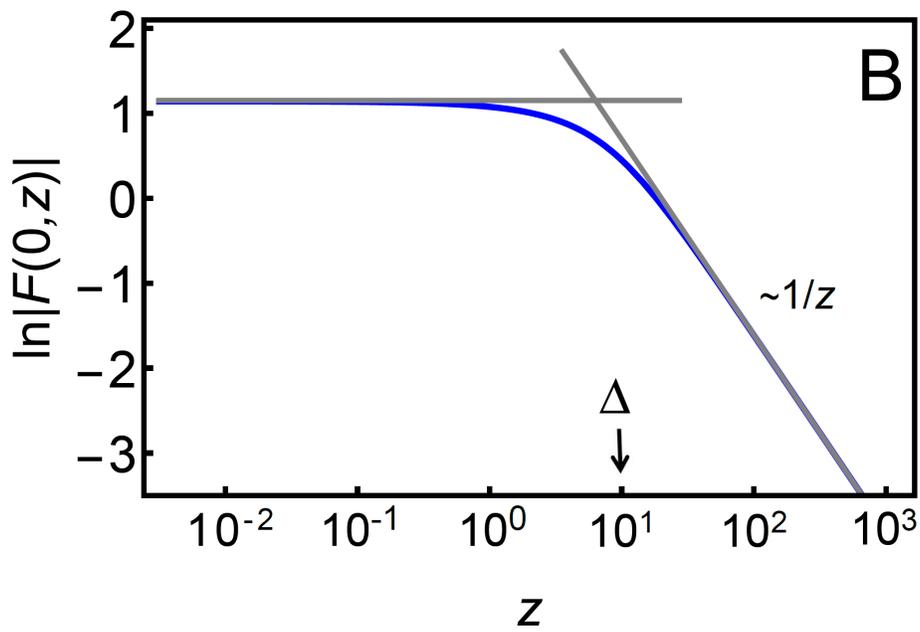

FIGURE 3

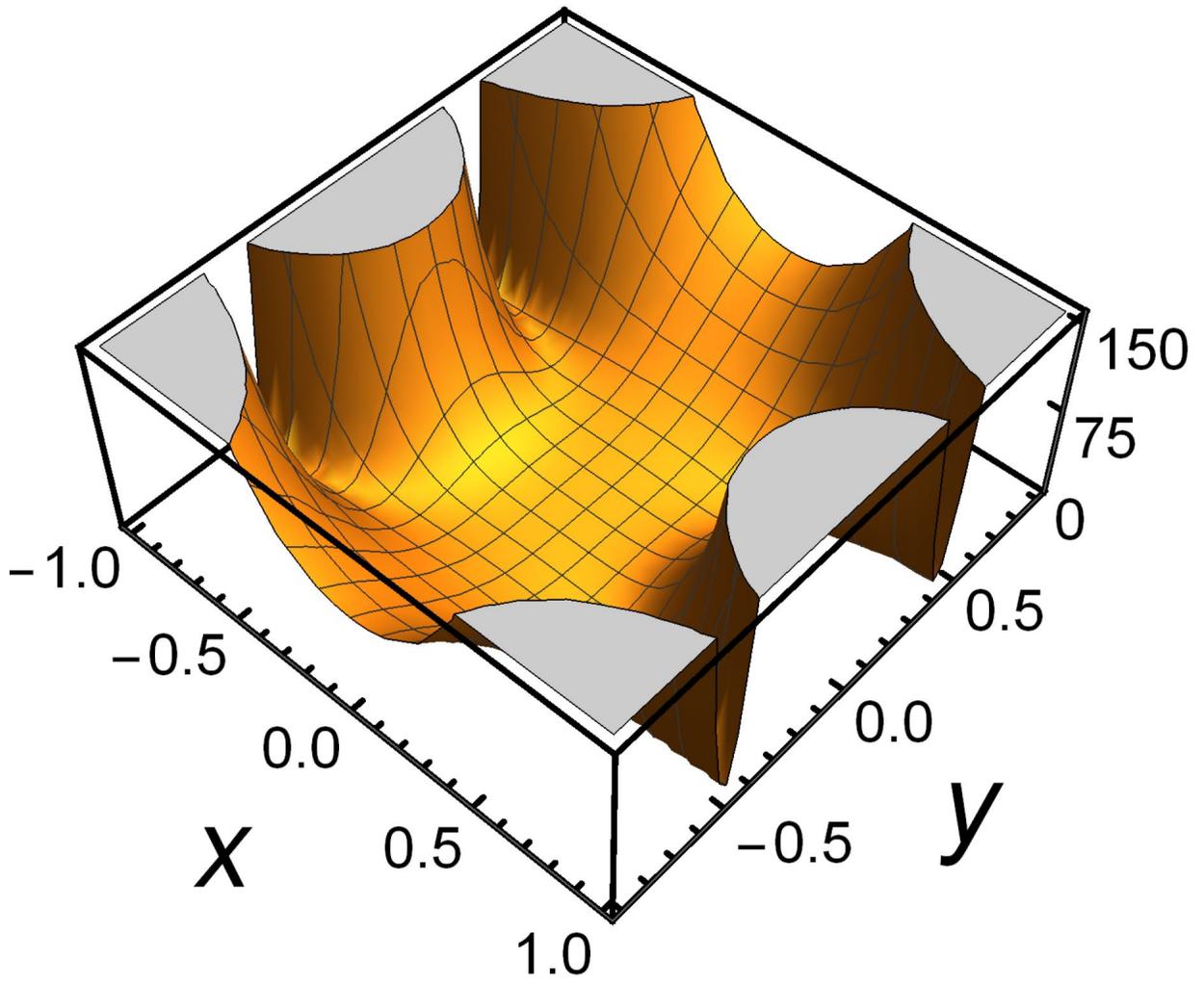

FIGURE 4